\documentclass{article}
\usepackage{amssymb}
\usepackage{amsfonts}
\usepackage{amsmath}
\usepackage{epsfig}
\usepackage{graphicx}

\begin{document}

\title{Quantum control of "quantum triple collisions" in a maximally
symmetric three-body Coulomb problem}
\author{R. Vilela Mendes\thanks{%
e-mail: rvilela.mendes@gmail.com, rvmendes@ciencias.ulisboa.pt;
https://label2.tecnico.ulisboa.pt/vilela/} \\
\textit{CMAFcIO, Universidade de Lisboa, }\\
\textit{\ C6 - Campo Grande, 1749-016 Lisboa}}
\date{ }
\maketitle

\begin{abstract}
In Coulomb 3-body problems, configurations of close proximity of the
particles are classically unstable. In confined systems they might however
exist as excited quantum states. Quantum control of such states by time
changing electromagnetic fields is discussed.
\end{abstract}

Availability of laser pulses of designed shape and very short time scales
provides a tool to control molecular dynamics. Quantum control applications
range from multi-photon excitations to direct control of chemical reactions
and to many diverse designs in quantum information \cite{Rabitz1} \cite%
{BOOK1} \cite{Koch} \cite{PRX}. By quantum control one might also be able to
excite exotic quantum states, in particular in confined systems \cite%
{MolContainer}. One type of such states are the \textit{scar }states \cite%
{Heller} \cite{Berry} \cite{VilelaSS1} which correspond to classically
unstable configurations but that may appear as well defined states in the
quantum spectrum.

This paper will be concerned with a $3$-body Coulomb problem of two
positively charged particles of mass $M$ and charge $Ze$ and a negatively
charged one of mass $m$ and charge $qe$. Let $\widetilde{R}$ be the
separation of the positive particles, $\widetilde{\rho }$ the distance of
the negative particle to one of the positive ones and $\psi \left( 
\widetilde{R},\widetilde{\rho }\right) $ the system wave function. The
question to be addressed is whether there are excited states for which $\psi
\left( 0,0\right) \neq 0$. Such states will be called "\textit{quantum
triple collisions}" or . As a possible practical application one addresses
the question of how to counter the Coulomb barrier of the $2$-body problem
by quantum control in the $3$-body problem. For one motivation to study this
problem refer to \cite{LConf-HybFus}. But even if the reader is uninterested
or skeptical about this motivation, the fact is that the problem of exciting
high-lying states by quantum control is interesting in its own right. Notice
that in classical mechanics triple collisions are singular points beyond
which the time evolution cannot be defined. Therefore there being no
classical periodic orbit corresponding to a triple collision, "quantum
triple collisions" are not, strictly speaking, scar states.

The central question in this paper, is the existence or non-existence of the
"quantum triple collision" states in the three-body Coulomb problem. In the
case of two heavy ($M$) and one light ($m$) particle, it is more or less
obvious, from kinetic barrier considerations, that such states, if they
exist, should be relatively high in the spectrum. Therefore to keep the
computational requirements reasonable, when the characterization of a large
number of states is desired, simplifications have to be introduced. Let $%
\overrightarrow{x}_{1}$, $\overrightarrow{x}_{2}$, $\overrightarrow{x}_{3}$
and $\overrightarrow{p}_{1}$, $\overrightarrow{p}_{2}$, $\overrightarrow{p}%
_{3}$ be the coordinates and momenta of the three particles. The first
simplification will be to concentrate on the dynamics of the two relative
coordinates ($\overrightarrow{x}_{1}-\overrightarrow{x}_{2}$,$%
\overrightarrow{x}_{1}-\overrightarrow{x}_{3}$) and the other to study only
maximally symmetric states. Once an energy level is obtained in this setting
what is the shift in energy as compared to the laboratory frame?

In the non-relativistic approximation the Coulomb potential only depends on
the modulus of the relative distances, therefore the correction originates
from the kinetic part and is%
\begin{equation*}
-\frac{\hbar ^{2}}{M}\overrightarrow{\nabla }_{1}\bullet \left( 
\overrightarrow{\nabla }_{2}+\overrightarrow{\nabla }_{3}-\frac{m}{2M}%
\overrightarrow{\nabla }_{1}\right)
\end{equation*}%
applied to the wave function. The first two terms are expected to be small
for maximal rotationally symmetric states, because they involve different
angle coordinates and the last one is suppressed by the ratio $m/M$.
Therefore even the energy levels and energy differences that are obtained
are probably not very different from those in the laboratory frame.

The full system has $9$ spatial degrees of freedom. However in the setting
described above and because the main purpose is to exhibit the existence of 
\textit{quantum triple collision states} it suffices to show their existence
in a subset of maximally symmetric states. Then the number of degrees of
freedom may be reduced to two. Take one of the positively charged particles
as the origin and use spherical coordinates for the other two particles. At
this stage the Hilbert space measure is%
\begin{equation}
d\nu =\widetilde{R}^{2}d\widetilde{R}d\Omega _{+}\widetilde{\rho }^{2}d%
\widetilde{\rho }d\left( \cos \theta \right) d\varphi  \label{1}
\end{equation}%
$\left( \widetilde{R},\Omega _{+}\right) $ being the coordinates of the
second positively charged particle and $\left( \widetilde{\rho },\theta
,\varphi \right) $ those of the negatively charged one. The Hamiltonian is%
\begin{equation}
\widetilde{H}=-\frac{\hbar ^{2}}{2M}\frac{1}{\widetilde{R}^{2}}\frac{%
\partial }{\partial \widetilde{R}}\left( \widetilde{R}^{2}\frac{\partial }{%
\partial \widetilde{R}}\right) -\frac{\hbar ^{2}}{2m}\frac{1}{\widetilde{%
\rho }^{2}}\frac{\partial }{\partial \widetilde{\rho }}\left( \widetilde{%
\rho }^{2}\frac{\partial }{\partial \widetilde{\rho }}\right) +V\left( 
\widetilde{R},\widetilde{\rho },\theta \right)  \label{2}
\end{equation}%
\begin{equation}
V\left( \widetilde{R},\widetilde{\rho },\theta \right) =\frac{Z^{2}e^{2}}{%
4\pi \varepsilon _{0}}\frac{1}{\widetilde{R}}-\frac{Zqe^{2}}{4\pi
\varepsilon _{0}}\left\{ \frac{1}{\widetilde{\rho }}+\frac{1}{\sqrt{%
\widetilde{R}^{2}+\widetilde{\rho }^{2}-2\widetilde{R}\widetilde{\rho }\cos
\theta }}\right\}  \label{3}
\end{equation}%
Let%
\begin{eqnarray}
\mu &=&\frac{m}{M}  \notag \\
G^{2} &=&\frac{Zme^{2}}{2\pi \varepsilon _{0}\hbar ^{2}}  \label{4}
\end{eqnarray}%
and redefine%
\begin{equation}
R=G^{2}\widetilde{R};\;\rho =G^{2}\widetilde{\rho };\;H=\frac{2m}{\hbar
^{2}G^{4}}\widetilde{H}  \label{5}
\end{equation}%
$\mu ,R,\rho $ and $H$ being dimensionless quantities, the results may
easily be used both for molecular and nuclear environments. For maximally
symmetric states, one may integrate over the angle variables obtaining%
\begin{equation}
H=\frac{2m}{\hbar ^{2}G^{4}}\widetilde{H}=-\mu \frac{1}{R^{2}}\frac{\partial 
}{\partial R}\left( R^{2}\frac{\partial }{\partial R}\right) -\frac{1}{\rho
^{2}}\frac{\partial }{\partial \rho }\left( \rho ^{2}\frac{\partial }{%
\partial \rho }\right) +\frac{Z}{R}-\frac{q}{\rho }-q\left\{ \frac{\chi
\left( R-\rho \right) }{R}+\frac{\mathcal{\chi }\left( \rho -R\right) }{\rho 
}\right\} ,  \label{6}
\end{equation}%
$\chi $ being the Heaviside function\footnote{$\mathcal{\chi }\left(
x\right) =1$ for $x\geq 0$, $\mathcal{\chi }\left( x\right) =0$ for $x<0$}.
The maximally symmetric system becomes a two degrees of freedom system with
integration measure 
\begin{equation}
d\nu =R^{2}dR^{2}\rho ^{2}d\rho  \label{7}
\end{equation}%
From (\ref{6}) one sees that, in spite of the Coulomb barrier between the
positive charges ($\frac{Z}{R}$), the effective potential becomes attractive
in the region $\rho <R$ if $\rho <\frac{q}{Z-q}R$. Given an eigenstate $\psi
\left( R,\rho \right) $ of $H$, the quantum probability for a two-body
collision of the positively charged particles is proportional to%
\begin{equation}
I_{2}=\int d\rho \rho ^{2}\left\vert \psi \left( 0,\rho \right) \right\vert
^{2}  \label{8}
\end{equation}%
and, as defined above, there is a quantum triple-collision if $\psi \left(
0,0\right) \neq 0$.

\begin{figure}[htb]
\centering
\includegraphics[width=0.75\textwidth]{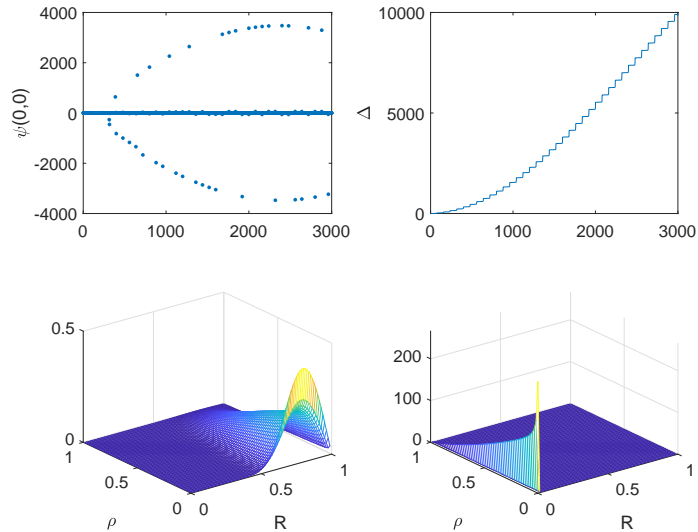}
\caption{$\protect\psi\left(0,0\right)$, the energy difference $\Delta$
between the ground and the excited states and the wave functions of the
ground state and the first quantum triple collision state}
\label{control_fig1}
\end{figure}

A difficulty on the way to a rigorous solution to this problem is the fact
that the potential is singular at the $R=\rho =0$ point. In an actual
physical system this point could never be reached because of the finite
dimensions of the particles. Therefore a reasonable approximation that
avoids the singularity problem is to compute the numerical solution of the
spectrum in a grid that does not contain the $R=\rho =0$ point, with the
average of $\psi $ on the smallest square around the origin standing for $%
\psi \left( 0,0\right) $. Because of the Coulomb barrier and the kinematical
cost of localization, it is to be expected that the quantum triple collision
states, if they exist, will be relatively high\ in the spectrum. Therefore
to compute them one needs a method that involves very many basis states. A
simple way to fulfill such a requirement is to represent the operator $H$ in
a fine grid of points in a box of size $\left[ 0,L\right] ^{2}$ \footnote{%
Because one is using spherical coordinates this box size corresponds roughly
to a lattice volume $\frac{4}{3}\pi L^{3}$.} and diagonalize the resulting
matrix. Fig.\ref{control_fig1} shows the results of such calculation for $%
\mu =2.7\times 10^{-4}$\footnote{%
This value corresponds roughly to the ratio of the electron and deuteron
masses. Using this value emphasizes the fact that quantum triple collision
states do exist even for small values of $\mu $, in spite of the kinetic
penalty associated to the small mass particle. For larger values of $\mu $
these states also exist, lower in the spectrum.}. The upper left panel is
the value of $\psi \left( 0,0\right) $ along the spectrum. One sees that for
all the lower part of the spectrum this is a vanishing value, although for
high excitation values there are many quantum triple collision states. These
states are many, but still somewhat exceptional in the whole set.

The right upper panel shows the \textit{energy difference }$\Delta $ between
the ground state $\psi _{0}$ and the excited states (in $H$ units) and the
two lower panels show respectively the wave functions of the ground state $%
\psi _{0}$ and of the first quantum triple collision state $\psi _{1}^{\ast
} $.

The objective now is to assess the possibility of carrying the system from
the ground state $\psi _{0}$ to the state $\psi _{1}^{\ast }$. The most
effective way for coherently controlling the evolution of a quantum system
is through the interaction between the system and an electromagnetic field
whose spectral content and temporal profile may be altered throughout the
process. The evolution equation would be%
\begin{equation}
i\hslash \partial _{t}\psi \left( t\right) =\left( H-\epsilon \left(
t\right) H_{1}\right) \psi \left( t\right)  \label{9}
\end{equation}%
where $H$ is the original Hamiltonian, $H_{1}$ the control operator and $%
\epsilon \left( t\right) $ the time varying control intensity. A well
established technique of optimal control \cite{Rabitz-1992} \cite%
{Tannor-1992} \cite{Turicini-2003} defines a function $F$ , to be minimized,
which contains both the objective goal and all the desired control
constraints, among them the equation of motion (\ref{9}). The constraints
are made independent by the introduction of Lagrangian multiplier fields and
the optimal control intensity $\epsilon \left( t\right) $ is obtained by
iterative forward integration of (\ref{9}) and backward integration of the
Lagrange multiplier equations. This method allows the introduction of
arbitrary control constraints, in particular the fluency $\int \epsilon
\left( t\right) ^{2}dt$ of the control field. An alternative local field
approach, which will be used here, defines a Lyapunov function \cite%
{Lyapunov}%
\begin{equation}
M\left( t\right) =\left( \psi \left( t\right) ,\psi _{1}^{\ast }\right)
\left( \psi _{1}^{\ast },\psi \left( t\right) \right)  \label{10}
\end{equation}%
and chooses $\epsilon \left( t\right) $ during the evolution, with initial
condition $\psi _{0}$, to insure that $\frac{d}{dt}M\geq 0$. Because%
\begin{equation}
\frac{dM\left( t\right) }{dt}=-\epsilon \left( t\right) \frac{2}{\hslash }%
\mathnormal{Im}\left\{ \left( \psi \left( t\right) ,\psi _{1}^{\ast }\right)
\left( \psi _{1}^{\ast },H_{1}\psi \left( t\right) \right) \right\}
\label{11}
\end{equation}%
the condition $\frac{d}{dt}M\geq 0$ is satisfied if%
\begin{equation}
\epsilon \left( t\right) =-\alpha \mathnormal{Im}\left\{ \left( \psi \left(
t\right) ,\psi _{1}^{\ast }\right) \left( \psi _{1}^{\ast },H_{1}\psi \left(
t\right) \right) \right\}  \label{12}
\end{equation}%
$\alpha $ a positive constant.

Even for a system contained in a box, the Hilbert space of solutions of the
equation (\ref{9}) is infinite-dimensional and it is known that full quantum
controllability in infinite dimensions is a delicate problem \cite{Turicini}
requiring non-Lie algebraic operators or approximations thereof \cite%
{Karwowski} \cite{Manko}. In this case however one deals with a simpler
problem of controllability between two states in a discrete spectrum. It is
known that in this case a necessary condition \cite{Chambrion} is
transitivity of the operator $H_{1}$. Hence the first thing to check is the
availability of $H_{1}$ operators that are transitive between these two
states, in the sense that there is an iteration $H_{1}^{n}$ of the operator
with non-vanishing matrix elements between the two states.

In the dipole approximation the interaction of the charged particles with
the electric field of a laser pulse takes place through the dipole operator,
namely%
\begin{equation}
D=\left( R-\rho \right) \cdot E  \label{13}
\end{equation}%
with $Z=q=1$, $E$ the electrical field and all constants included in $%
\epsilon \left( t\right) $. To obtain the effect of this operator on the
maximally symmetric states one integrates over all angles obtaining%
\begin{equation}
H_{1}^{(E)}=\frac{1}{16R\rho }\left\{ \left( R+\rho \right) ^{3}-\left\vert
R-\rho \right\vert ^{3}\right\}  \label{14}
\end{equation}

\begin{figure}[htb]
\centering
\includegraphics[width=0.8\textwidth]{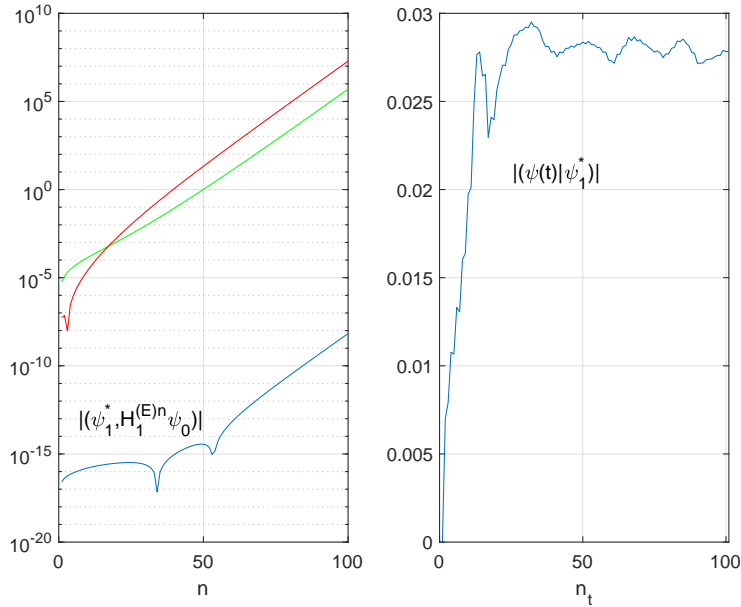}
\caption{$\left\vert \left( \protect\psi _{1}^{\ast },H_{1}^{(E)^{n}}\protect%
\psi _{0}\right) \right\vert $ compared with similar matrix elements
replacing $\protect\psi _{1}^{\ast }$ by two states with $\protect\psi %
(0,0)=0$ and a control attempt with the dipole operator. $n-_t$ is the
number of control steps with $dt=0.05$ }
\label{control_fig2}
\end{figure}

The transitivity of this operator is found by computing $\left\vert \left(
\psi _{1}^{\ast },H_{1}^{(E)^{n}}\psi _{0}\right) \right\vert $ for
successive powers of the operator. The result is shown in the left-hand
panel of Fig.\ref{control_fig2} where this value is compared with the
corresponding matrix element with $\psi _{1}^{\ast }$ replaced by two other
randomly chosen states for which $\psi \left( 0,0\right) =0$. This shows
that, starting from the ground state, the quantum triple collision state is
not controllable with this operator. This is confirmed in the right-hand
panel where a control attempt is made using the Lyapunov method and
adjusting the field $\epsilon \left( t\right) $ intensity at each step by (%
\ref{12}). The time step is $dt=0.05$ and the exponential of the operator is
computed at every step to improve the precision. One sees that the overlap $%
\left\vert \left( \psi _{0},\psi _{1}^{\ast }\right) \right\vert $ always
remains at the level of the numerical round-off. The uncontrollability by
the dipole operator is in fact to be expected because in the quantum triple
collision state $R$ and $\rho $ are expected to be small, even for states
that are not maximally symmetric. In the maximally symmetric case, studied
here, one sees from Fig.\ref{control_fig1} that in the $\psi _{1}^{\ast }$
state $R\approx 0$ and therefore the operator $H_{1}^{(E)}$ in (\ref{14})
vanishes.

For a controlling alternative one considers the interaction with a magnetic
field $B$. This interaction has two different terms, the paramagnetic and
the diamagnetic, both arising from the substitution $p\rightarrow p-eA\left(
x,t\right) $ . The paramagnetic term may be written%
\begin{equation*}
-\frac{e}{2m}L\cdot B
\end{equation*}%
and therefore, being proportional to the orbital angular momentum $L$, it
vanishes for a maximally symmetric state. The diamagnetic term is
proportional to%
\begin{equation*}
\rho ^{2}\left( B_{z}^{2}+B_{y}^{2}\right) +Z\mu R^{2}
\end{equation*}

\begin{figure}[htb]
\centering
\includegraphics[width=0.8\textwidth]{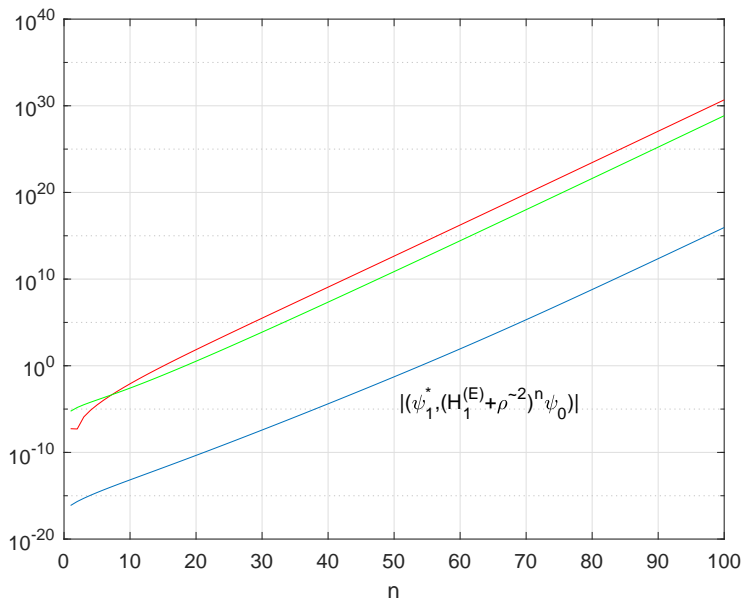}
\caption{$\left\vert \left( \protect\psi _{1}^{\ast },\left( H_{1}^{(E)}+%
\protect\rho^{2}\right)^{n}\protect\psi _{0}\right) \right\vert $ compared
with similar matrix elements replacing $\protect\psi _{1}^{\ast }$ by two
states with $\protect\psi (0,0)=0$}
\label{control_fig3}
\end{figure}

If $M\gg m$, $\mu $ is very small and will be mostly the operator $\rho
^{2}\ $\ that might have a controlling effect. In Fig.\ref{control_fig3} one
shows the successive values of $\left\vert \left( \psi _{1}^{\ast },\left(
H_{1}^{(E)}+\rho ^{2}\right) ^{n}\psi _{0}\right) \right\vert $. Ones sees
that in this case the matrix element becomes large, although not so large as
the corresponding matrix elements for the same two reference states as used
in Fig.\ref{control_fig2} . Some degree of controllability is confirmed by
using again the Lyapunov method now with the operator%
\begin{equation*}
H_{1}^{(M)}=H_{1}^{(E)}+\rho ^{2}
\end{equation*}

\begin{figure}[htb]
\centering
\includegraphics[width=0.8\textwidth]{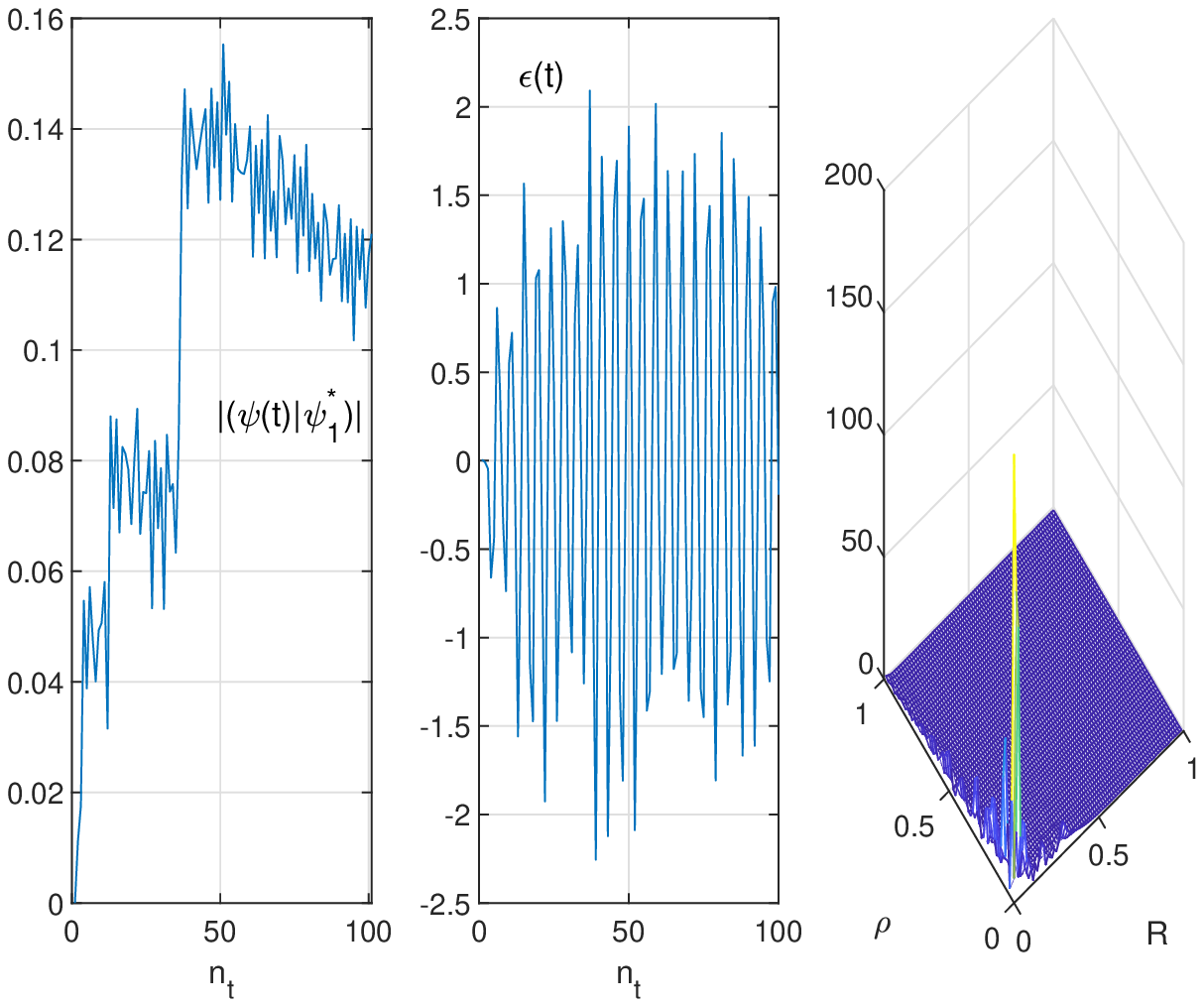}
\caption{Control with the $H_{1}^{(M)}$ operator. The overlap $\left\vert
\left( \protect\psi \left( t\right) ,\protect\psi _{1}^{\ast }\right)
\right\vert $, the control intensity $\protect\epsilon \left( t\right) $ and
the wave function $\protect\psi \left( t\right) $ after $100$ iterations}
\label{control_fig4}
\end{figure}

The result of the numerical calculation is shown in Fig.\ref{control_fig4}
where one sees that the overlap indeed grows rapidly on the first four
iterations then settling around $12\%$. The left-hand panel shows the
overlap $\left\vert \left( \psi \left( t\right) ,\psi _{1}^{\ast }\right)
\right\vert $, the middle one the control intensity $\epsilon \left(
t\right) $ and the right-hand panel the wave function $\psi \left( t\right) $
after $100$ iterations. Although the controlled wave function is very close
to a quantum triple collision situation, the overlap is still small because,
as seen in right-hand panel, the coincidence with the objective function is
mostly in the region of small $\rho $ and $R$ where the integration measure (%
\ref{7}) is small.

In conclusion: A Coulomb system of two positive and one negative charge
confined in a box has quantum triple collision states. These states are high
excited states in the spectrum. They are many but still exceptional in a
"sea" of states with $\psi \left( 0,0\right) =0$. Quantum control from the
ground state, by the dipole operator, is not possible in maximally symmetric
states and also expected to be inefficient for non-symmetric states.
However, it seems possible using time-varying magnetic fields. Magnetic
control might even be more efficient for non-symmetric states because of the
action of the paramagnetic term.

Notice however that the non-controllability result with the dipole operator
and the electric field refers only to exact controllability, which arises
from the non-transitivity of the $H_{1}^{(E)}$ operator. What one observes,
for example in the control experiment reported in the right-hand panel of
Fig.\ref{control_fig2}, is that the controlled wave function $\psi \left(
t\right) $ converges to states of close proximity ($R$ small), nevertheless
with negligible overlap with the objective function $\psi _{1}^{\ast }$.

In an actual 3-body Coulomb system confined in a solid lattice, accurate
calculation of the energy levels will be difficult, because it is strongly
influenced by the solid state environment. Therefore to have success in the
use of quantum triple collisions to induce molecular or nuclear reactions,
some experimental automatic learning process as in \cite{Rabitz2} is
recommended.

\end{document}